\documentclass[pra,twocolumn,showpacs,preprintnumbers,fleqn,superscriptaddress]{revtex4-1}
\usepackage{amsmath,amsfonts,amssymb}
\usepackage{graphicx}

\usepackage{color}\definecolor{gray}{rgb}{0.5,0.5,0.5}

\begin{document}

\title{High-order harmonic generation from gapped graphene: perturbative response and transition to non-perturbative regime}

\author{Darko Dimitrovski}
\affiliation{Department of Physics and Nanotechnology, Aalborg
University, Skjernvej 4A 9220 Aalborg East, Denmark}
\email{dd@nano.aau.dk}
\author{Lars Bojer Madsen}
\affiliation{Department of Physics and Astronomy, Aarhus
University, Ny Munkegade 120, 8000 Aarhus C, Denmark}
\author{Thomas Garm Pedersen}
\affiliation{Department of Physics and Nanotechnology, Aalborg
University, Skjernvej 4A 9220 Aalborg East, Denmark}

\date{\today} 

\begin{abstract}

We consider the interaction of gapped graphene in the two-band
approximation using an explicit time-dependent approach. In
addition to the full high-order harmonic generation (HHG)
spectrum, we also obtain the perturbative harmonic response using
the time-dependent method at photon energies covering all the
significant features in the responses. The transition from the
perturbative to the fully non-perturbative regime of HHG at these
photon energies is studied in detail.

\end{abstract}



\def\eps{\varepsilon}
\def\i{\mathrm{i}}
\def\e#1{\mathrm{e}^{#1}}
\def\Ejo{E_{j0}}
\def\Eoj{E_{0j}}
\def\dE{\Delta}

\def\fs#1{\bar{#1}}\def\fs#1{#1_{\rm fs}}
\def\wfs{\fs{\omega}}\def\tfs{\fs{\tau}}
\def\vFfs{\fs{\mathbf{F}}}\def\Ffs{\fs{F}}\def\ffs{\fs{f}}

\def\as#1{\widetilde{#1}}
\def\was{\as{\omega}}\def\as#1{#1_{\rm as}}
\def\vFas{\as{\mathbf{F}}}\def\Fas{\as{F}}\def\fas{\as{f}}
\def\Tas{\as{T}}\def\tas{\as{\tau}}

\def\Eas{\as{E}}

\def\Epmi{\Eas{\pm} \wfs}
\def\Empi{\Eas{\mp} \wfs}
\def\Emi{\Eas{-} \wfs}
\def\Epi{\Eas{+} \wfs}
\def\Epmii{\Eas{\pm} \wfs}
\def\Empii{\Eas{\mp} \wfs}


\def\Iint#1#2{I_{#1}(#2)}\def\bIint#1#2{\bar{I}_{#1}(#2)}
\def\Jint#1#2{J_{#1}(#2)}\def\Lint#1#2{\bar{J}_{#1}(#2)}
\def\Ixint#1#2{I^{*}_{#1}(#2)}
\def\Jxint#1#2{J^{*}_{#1}(#2)}

\def\p\A#1{P(#1)}

\def\a{\alpha}
\def\b{\beta}
\def\c{\gamma}
\def\d{\delta}
\def\ps#1{p_{#1}}

\def\E{{\cal E}}
\def\A{{\cal A}}

\maketitle

\section{Introduction}

The interaction of strong lasers with solids has been studied
since the early days of strong-field physics \cite{Keldysh1965}.
Recently, due to the development of short, strong laser pulses
with controlled waveforms \cite{KrauszStockman_Review2014}, it has
become relevant to consider the response of such materials to
strong laser pulses with respect to the transferred charge
\cite{Krausz2013_1,Stockman2015}, and the generated harmonic
radiation \cite{Ghimire2011,BrabecPRL2014}. As the pulses used in
these studies are strong and short, they come almost exclusively
from (near) infrared sources.

A material of special interest is graphene. The properties of
graphene, such as its stability and the huge mobility of carriers,
promise a plethora of nanoscale electronic applications
\cite{Novoselov2007}. Concerning harmonic radiation by strong
laser fields, in the past HHG in graphene was considered by
directly applying the strong-field approximation
\cite{Keldysh1965,Lewenstein1994} for graphene described on the
level of molecular orbitals \cite{Jan-Petter2013,Chinese2016}. HHG
in graphene was also considered performing time domain
calculations that took into account the inter- and intraband
dynamics for THz pulses and in the Dirac approximation
\cite{AN2014,Ishikawa2010}, and calculations investigating
multiphoton resonant excitation
\cite{Avetissian2012,Avetissian2012Nano,Avetissian2013}. Another
very active area of research is the investigation of the third
harmonic generation in graphene, for recent results, see e.g.,
\cite{Cheng2015,Rostami2016}. Graphene is, however, a semimetal
with a zero band gap, and that limits the possible applications in
electronic and optoelectronic devices. Fortunately a class of
materials, termed gapped graphene, based on or similar to graphene
was developed using various techniques
\cite{Zhou2007,Castro2007,Li2008,TGP2008_1,TGP2008_2}. Gapped
graphene can be described within the  two-band tight-binding
approximation \cite{TGP2009}. This enabled extensive theoretical
studies of optical response of this system including the linear
\cite{TGP2009,Pyatkovskiy} and beyond linear response
\cite{arXiv2016}, second harmonic generation \cite{TGP2015}, third
harmonic response \cite{Jafari,Sipe2015} and magneto optics
\cite{pedersenx2}.

Of particular interest is the ability of the theory to identify
the breakdown of perturbation theory and to deal directly with the
explicit time-dependence of the pulse. Here we therefore consider
high-order harmonic generation spectra for gapped graphene from
the perturbative optical response and into the non-perturbative
regime. In particular, we consider the first, second and third
harmonic responses using a time-dependent approach and investigate
the breakdown of perturbation theory.

The paper is organized as follows. In the next section we present
the basic structure and the equations for the two-band model of
gapped graphene. In Sec. III we present the basics of the
interaction of a two-band system with light, including the way to
calculate high-order harmonic generation (HHG) spectra. In Sec. IV
we compare the harmonic response to the perturbative harmonic
response for gapped graphene. The transition from the perturbative
harmonic response to the non-perturbative HHG spectra is
considered in Sec. V, where we also consider the gap dependence.
We conclude in the last section. The expressions for the dipole
couplings and momentum matrix elements within and between the
bands of gapped graphene are given in the Appendices.

\section{Structure and basic equations}

\begin{figure}
\begin{center}
\includegraphics[width=0.3\textwidth]{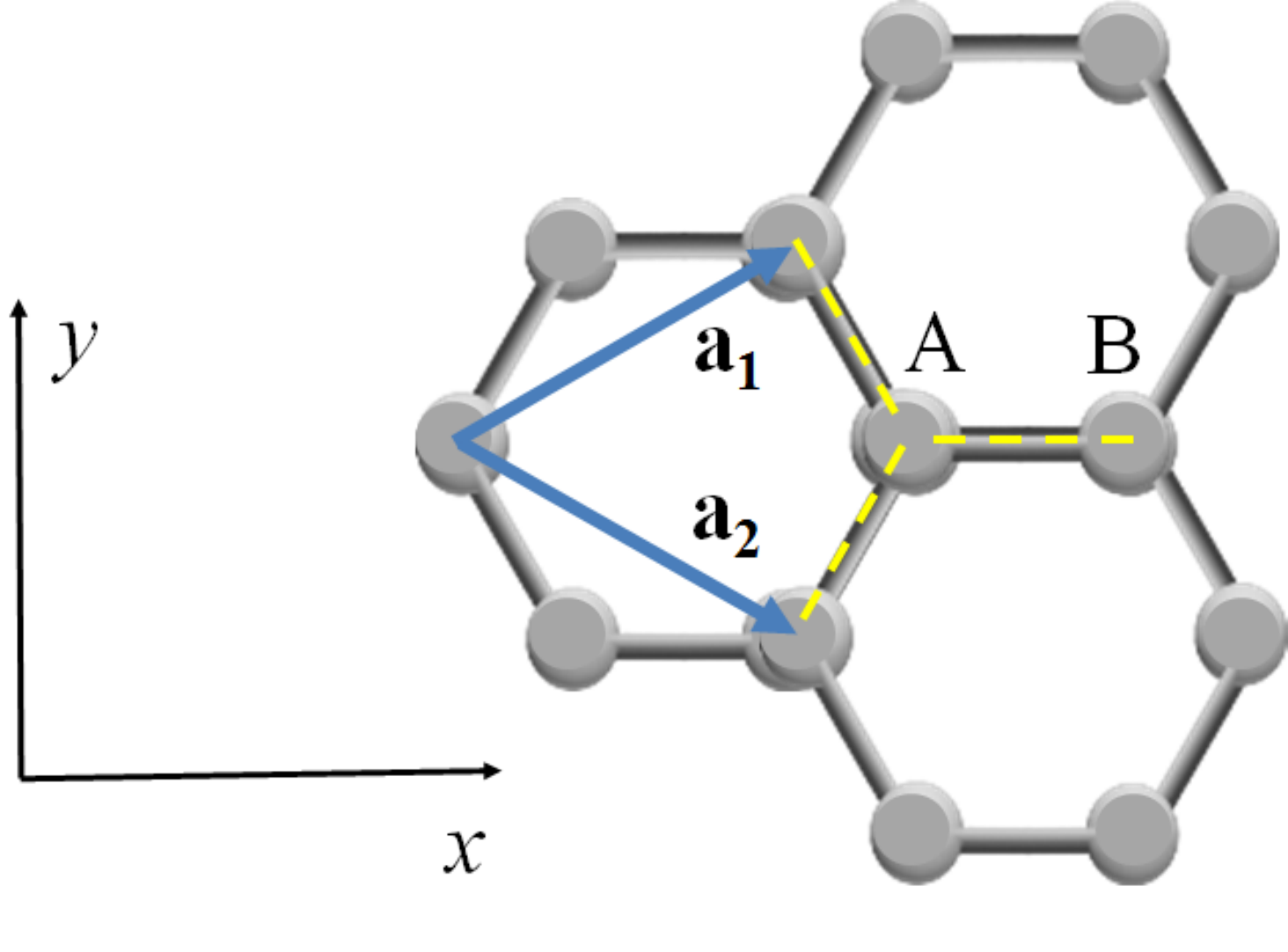}
\end{center}
\caption{The structure of graphene and gapped graphene in position
space. The elementary lattice vectors, $\mathbf{a}_1$ and
$\mathbf{a}_2$, as well as the atomic sites A and B (the two
inequivalent sublattices) are shown. The yellow dashed lines
denote the connections of A with its nearest neighbors. The $x$
and $y$ axes are also indicated.} \label{fig:orient}
\end{figure}

The structure of graphene, and also of gapped graphene, which is
identical in position space, is given in Fig. \ref{fig:orient}.
The elementary lattice vectors, shown in Fig. \ref{fig:orient},
are \cite{Wallace1947}
\begin{equation}
\textbf{a}_1 = \dfrac{a}{2} \left(
\begin{array}{c}
\sqrt{3}\\
1\\
\end{array}
\right), \quad \textbf{a}_2 = \dfrac{a}{2} \left(
\begin{array}{c}
\sqrt{3}\\
-1\\
\end{array}
\right) ,
\end{equation}
where $a=2.46$ \r{A} is the lattice constant.

To obtain the electronic band structure, we use the $p_z$ atomic
orbitals at the atomic sites A and B [Fig. \ref{fig:orient}], $
\left| p_z ( \textbf{r} - \textbf{R}_A ) \right> $ and $ \left|
p_z ( \textbf{r} - \textbf{R}_B ) \right> $. Then we form Bloch
wave functions $ \left| \alpha \right> = \frac{1}{\sqrt{N}}
\displaystyle\sum_{\textbf{R}} e^{i\textbf{k} \cdot (\textbf{R}_A
+ \textbf{R} )} \left| p_z ( \textbf{r} - ( \textbf{R}_A +
\textbf{R}) ) \right> $ and $ \left| \beta \right> =
\frac{1}{\sqrt{N}} \displaystyle\sum_{\textbf{R}} e^{i\textbf{k}
\cdot ( \textbf{R}_B + \textbf{R} )} \left| p_z ( \textbf{r} - (
\textbf{R}_B + \textbf{R}) ) \right> $, where $N \to \infty$ is
the number of unit cells, and the sum runs over the Bravais
lattice vectors $\mathbf{R}$, and $\mathbf{k}$ is the wave vector.

A band gap in graphene can be induced in several ways: graphene
grown on SiC substrate \cite{Zhou2007}, biasing a graphene bilayer
\cite{Castro2007}, sculpturing a graphene into nanoribons
\cite{Li2008}, or introducing a periodic array of circular holes
\cite{TGP2008_1,TGP2008_2}. In addition, systems like hexagonal
BN, where 2 carbon atoms in the unit cell are replaced by a BN
dimer, can be described using the same model as gapped graphene
with respect to the interaction with light \cite{BN_TB_2008}. Here
we focus on the class of gapped graphene where the inversion
symmetry is broken, such as graphene grown on the SiC surface and
the BN. For this type of gapped graphene, similarly to graphene
\cite{Wallace1947}, using $\left| \alpha \right>$ and $\left|
\beta \right>$ and assuming nearest-neighbor coupling, the
tight-binding Hamiltonian is obtained as \cite{TGP2009}

\begin{equation}
\hat{\textbf{H}}_0  =
  \begin{bmatrix}
    \frac{\Delta}{2} & - \gamma f ( \mathbf{k} )  \\
    &   \\
 - \gamma f^* ( \mathbf{k} )& -\frac{\Delta}{2}
  \end{bmatrix}, \label{eq:GG}
\end{equation}
where $\Delta$ is the energy gap, $ \gamma =  - \left< p_z (
\textbf{r} - \textbf{R}_A ) \left| \hat{H} \right| p_z (
\textbf{r} - \textbf{R}_B ) \right> \approx 3 \text{ eV}$ is the
hopping integral and
\begin{equation}
f ( \textbf{k} ) = \exp \left( i \dfrac{a k_x }{\sqrt{3}} \right)
+ 2 \exp \left( -i\dfrac{a k_x }{2 \sqrt{3}} \right) \cos \left(
\dfrac{a k_y }{2} \right) \label{eq:fk}
\end{equation}
comes from the geometry of the location of the nearest neighbors,
see Fig. \ref{fig:orient}, i.e., from the addition of factors of a
type $\exp( i \mathbf{k} \cdot ( \mathbf{R}_A - \mathbf{R}_B ))$.
Diagonalizing the Hamiltonian of Eq. \eqref{eq:GG} we recover the
valence band $ E_v ( \textbf{k} ) = - \sqrt{ \left(
\frac{\Delta}{2} \right)^2 + \gamma^2 \left| f ( \textbf{k} )
\right|^2  } $ and the conduction band $ E_c ( \textbf{k} ) =
\sqrt{ \left( \frac{\Delta}{2} \right)^2 + \gamma^2 \left| f (
\textbf{k} ) \right|^2  }$.

\section{Interaction with light in the dipole approximation for a two-band system}

For a two-band system, such as the one obtained using the
tight-binding approximation, the wavefunction can be written as
\begin{equation}
\Psi ( \textbf{r}, t) = \displaystyle\sum_{m=c,v}
\displaystyle\int_{\text{BZ}} a_m ( \textbf{k}, t )
\psi_{m,\textbf{k}} ( \textbf{r} ) d^3 \textbf{k}, \label{eq:wf}
\end{equation}
where BZ denotes that the integration is performed over the
Brillouin zone, $c$ and $v$ denote conduction and valence bands,
respectively, and
\begin{equation}
\psi_{m,\textbf{k}} ( \textbf{r} ) = u_{m\textbf{k}} ( \textbf{r}
) \exp \left( i \textbf{k} \cdot \textbf{r} \right)
\end{equation}
are the Bloch wave functions - eigenfunctions of the field-free
Hamiltonian $\hat{H}_0$, i.e., $  \hat{H}_0 \psi_{m,\textbf{k}} =
E_m ( \textbf{k} ) \psi_{m,\textbf{k}} ( \textbf{r} ) $. The
field-free Hamiltonian $\hat{H}_0$, which can describe any
two-band system, refers here to the Hamiltonian written in matrix
form in Eq. \eqref{eq:GG}.

When interacting with light, in the length gauge, $ \hat{H} (t) =
\hat{H}_0 + e \textbf{F} (t) \cdot \textbf{r}$, where $\mathbf{F}
(t)$ is the electric field of the laser and $e$ is the norm of the
electron charge. The $a_m$'s from Eq. \eqref{eq:wf} satisfy the
following equations of motion \cite{AS1995}
\begin{equation}
\dot{a}_{m} = \left( - \frac{ i}{\hbar} E_m ( \textbf{k} ) +
\frac{e}{\hbar} \textbf{F} (t) {\boldsymbol \nabla_{\textbf{k}} }
\right) a_m - i \frac{e}{\hbar} \textbf{F} (t) \cdot
\displaystyle\sum_{n} {\boldsymbol \xi}_{m n} ( \textbf{k} ) a_{n}
\label{eq:eqm_wf} \end{equation} where
\begin{equation}
{\boldsymbol \xi}_{m n} ( \textbf{k} ) = i \displaystyle\int
u^*_{m\textbf{k}} ( \textbf{r} ) {\boldsymbol \nabla}_{\textbf{k}}
u_{n\textbf{k}} ( \textbf{r} ) d^3 \textbf{r} , \label{eq:xi_def}
\end{equation}
$n,m\in (c,v)$, and where the dependence of $a_m$ and $a_n$ on $k$
and $t$ is omitted to ease notation. The explicit expressions for
the ${\boldsymbol \xi}$'s of Eq. \eqref{eq:xi_def} are given in
Appendix A.


The amplitude equations \eqref{eq:eqm_wf} do not readily allow
inclusion of decoherence and temperature effects. For this
purpose, we reformulate the equations of motion using the density
matrix to arrive at
\begin{equation}
 i \hbar \dfrac{ d \rho }{dt} = \left[ \hat{H}_0 + e \textbf{F} \cdot \textbf{r} , \rho \right],
\label{eq:eq_mt}
\end{equation}
where
\begin{equation}
 \left[ \hat{H}_0 , \rho \right]_{nm} = \left( E_n - E_m \right)
\rho_{nm} \label{eq:Hcom}
\end{equation}
and, following \cite{AS1995},
\begin{equation}
\left[ \textbf{r}^{(i)} , \rho \right]_{nm} = i {\boldsymbol
\nabla_{\textbf{k}}} \rho_{nm} + \rho_{nm} \left( {\boldsymbol
\xi}_{nn} - {\boldsymbol \xi}_{mm}  \right) . \label{eq:rcom}
\end{equation}
Inserting Eqs. \eqref{eq:Hcom} and \eqref{eq:rcom} in Eq.
\eqref{eq:eq_mt} and adding a term containing the decoherence
times, $\tau_1$ for $\rho_{cv}$ and $\tau_2$ for $n$, to introduce
a decay, we obtain the following coupled equations of motion
\begin{widetext}
\begin{equation} \dfrac{d \rho_{cv} ( \textbf{k} , t ) }{dt}   =  -i \omega_{cv} ( \textbf{k} )
\rho_{cv} ( \textbf{k} ,t )- i \dfrac{e}{\hbar} \textbf{F} (t)
\cdot {\boldsymbol \xi}_{cv} ( \textbf{k} ) n ( \textbf{k} , t)
  + \dfrac{e}{\hbar} \textbf{F} (t) \cdot
{\boldsymbol \nabla}_{\textbf{k}} \rho_{cv} ( \textbf{k} , t ) - i
\dfrac{e}{\hbar} \textbf{F} (t) \cdot \left( {\boldsymbol
\xi}_{cc} ( \textbf{k} ) - {\boldsymbol \xi}_{vv} ( \textbf{k} )
\right) \rho_{cv} ( \textbf{k} , t ) - \dfrac{\rho_{cv} (
\textbf{k} , t ) }{\tau_1} \label{eq:eq_mot1}
\end{equation} and
\begin{equation} \dfrac{dn ( \textbf{k} , t )}{dt}  =
2 i \dfrac{e}{\hbar} \mathbf{F} (t) \cdot \left( {\boldsymbol
\xi}_{cv} ( \textbf{k} ) \rho^*_{cv} ( \textbf{k} , t ) -
{\boldsymbol \xi}^*_{cv} ( \textbf{k} ) \rho_{cv} ( \textbf{k} , t
) \right) + \dfrac{e}{\hbar} \textbf{F} (t) \cdot {\boldsymbol
\nabla}_{\textbf{k}} n( \textbf{k} , t) - \dfrac{n( \textbf{k},t)
-(f_v (\textbf{k}) -f_c (\textbf{k}) ) }{\tau_2} ,
\label{eq:eq_mot2}
\end{equation}
\end{widetext}
where $  n = \rho_{vv} - \rho_{cc}$, $ \omega_{cv} = \dfrac{E_c -
E_v}{\hbar} $, and
\begin{equation}
f_{c/v} ( \textbf{k} , T ) = \left( 1+ \exp \left( \dfrac{E_{c/v}
( \textbf{k} ) }{k_B T} \right) \right)^{-1}  \end{equation} is
the Fermi-Dirac distribution for the conduction and valence band,
respectively. In the above equation, $k_B$ is Boltzmann's
constant, and $T$ is the temperature.

The equations of motion, \eqref{eq:eq_mot1} and
\eqref{eq:eq_mot2}, are solved with the initial conditions
$\rho_{cv} (\textbf{k} , -\infty ) = 0$ and $n (\textbf{k} ,
-\infty ) = f_v (\textbf{k},T) -f_c (\textbf{k},T)$. The numerical
approach for solving the above equations is based on Ref.
\cite{AN2014}: we use a $\textbf{k}$ grid and approximate the
gradients with balanced difference. The time propagation is
performed using an adaptive Runge-Kutta algorithm. As in Ref.
\cite{AN2014}, we use $T=10$ K throughout.

\subsection{Calculating the current and the HHG spectrum}

The current density, to be referred to simply as the current
throughout, is given by $\textbf{J} (t) = - 2 \dfrac{e }{m_e }
\text{Tr} \{ \hat{ \mathbf{p}} \rho \} $, where the factor of 2
stands for the spin multiplicity. The current is explicitly
calculated as
\begin{eqnarray}
\textbf{J} (t)  =   -  \dfrac{e }{ 2 \pi^2 m_e } \left[ \int d
\textbf{k} \left( \textbf{p}_{vc} (\textbf{k} ) \rho_{cv}
(\textbf{k} , t) + \textbf{p}_{cv} (\textbf{k} )  \rho_{vc} (
\textbf{k} , t) \right) \right. \nonumber \\  \left. + \int d
\textbf{k} \dfrac{1}{2} (\textbf{p}_{vv} ( \textbf{k} ) -
\textbf{p}_{cc} ( \textbf{k} ) ) n ( \textbf{k} , t ) )\right] .
\label{eq:current}
\end{eqnarray}
The first integral in the above equation represents the interband
current, while the second integral the intraband current. The
momentum matrix elements appearing in the expression for the
current are obtained as follows. The diagonal momentum matrix
elements are obtained as $  \textbf{p}_{nn} = \dfrac{m_e }{\hbar}
\dfrac{\partial E_n }{\partial \textbf{k}}$, while the
off-diagonal matrix elements can be obtained as either $
\textbf{p}_{nm} = \dfrac{m_e}{\hbar}  \left< n \textbf{k} \left|
\dfrac{\partial \hat{H} }{\partial \textbf{k}}  \right| m
\textbf{k} \right>$ \cite{Pedersen2003,Pedersen2001R} or $
\textbf{p}_{nm} = i m_e \omega_{nm} {\boldsymbol \xi}_{nm}$
\cite{AS1995}. The explicit expressions for the momentum matrix
elements are given in Appendix B.

The harmonic spectrum is obtained as $ \left| \textbf{\cal{j}} (
\Omega ) \right|^2 $, where $  \textbf{\cal{j}} ( \Omega ) = {\cal
F} \left\{ \textbf{J} (t) \right\}$ is the Fourier transform of
the current. In practice we consider the component of the current
along one direction, in our case the $x$-component of the current
[Fig. \ref{fig:orient}], and present the discrete Fourier
transform of the current
\begin{equation} j(\Omega ) = \sum_{k=0}^{N_p -1} J_k \exp (i
\Omega t_k ) ,\label{eq:DFT}
\end{equation}
where $N_p$ is the number of points for current samples ($J_k$)
and time samples ($t_k$). We use a laser pulse, defined by the
electric field vector
\begin{eqnarray} \label{eq:field}
\mathbf{F} (t) = \mathbf{F}_0 \exp \left[ - \left( \frac{t-MT_p
/2}{MT_p /6} \right)^2 \right] \sin \left( \frac{ 2 \pi}{T_p} t
\right) \\ \text{ for }t\in[0,MT_p] , \nonumber
\end{eqnarray}
where $F_0= | \textbf{F}_0 |$ is the peak electric field strength,
$T_p =2 \pi / \omega$ is the period of the field, with $\omega$
the driving frequency, and $M$ is the number of the field cycles.
The exponential factor in Eq. \eqref{eq:field} describes the
envelope and the sinusoidal factor the carrier of the pulse. We
express the peak field strength in atomic units (a.u.) - 1 a.u. of
field strength is $5.142 \times 10^{11}$ V/m. The Fourier
transform of the field scaled by its duration ($M T_p $), in the
limit of infinitely large pulse ($M \to \infty$) and for each
$\omega$ is proportional to a $\delta$ function in Fourier space.
This scaling is exploited for the current - the expression
\eqref{eq:DFT} does not depend on the pulse duration explicitly
therefore in the limit of infinitely long pulses \eqref{eq:DFT} is
proportional to the Fourier transform of the current caused by an
infinite periodic pulse. Finally, in this way, the discrete
Fourier transform of the current \eqref{eq:DFT} has the same
dimension as the current.

\section{Perturbative harmonic response}

We consider the harmonic responses of gapped graphene,
traditionally investigated using frequency-domain methods
\cite{Boyd}, using explicitly time-dependent methods. We do this
to test our numerical solution and to investigate the breakdown of
perturbation theory.

For illustrative calculations, capturing generic effects in gapped
graphene, we consider a gap of 1 eV. To ensure that well-defined
harmonic peaks appear we perform calculations using pulses
described by Eq. \eqref{eq:field} with $M=48$ cycles. Next, to
stabilize the numerical calculations and ensure rapid convergence
we choose a relatively small value of the decoherence time $\tau_1
= \tau_2 = 5$ fs. We orient the field along the $x$-axis [see Fig.
\ref{fig:orient}] so that both odd and even harmonics appear. To
extract the first and the second harmonic response from the
numerical calculations we first obtain the full harmonic spectrum
for a fixed driving frequency $\omega$ and then select only the
value at the first and second harmonic, and repeat the procedure,
changing $\omega$ in small steps to ensure that all the features
in the harmonic responses are captured.

\begin{figure}
\begin{center}
\includegraphics[width=0.49\textwidth]{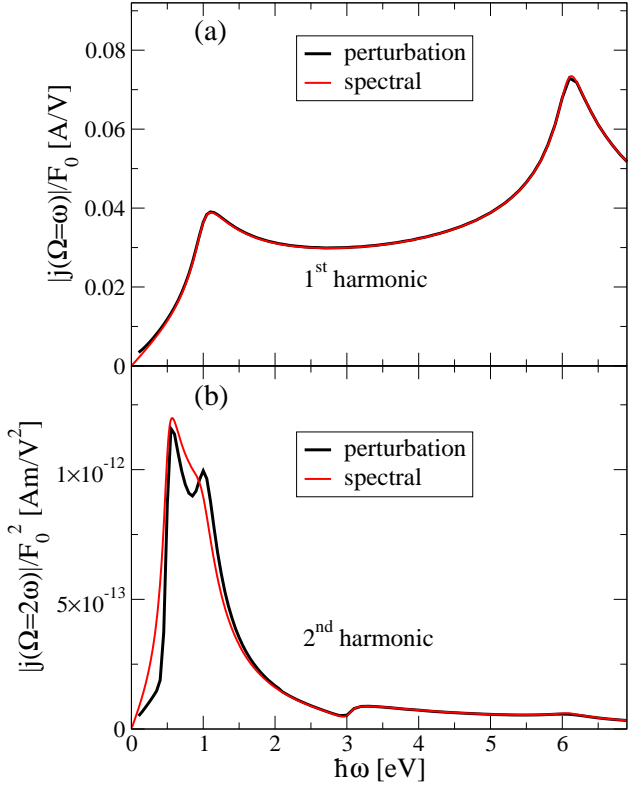}
\end{center}
\caption{The absolute value of the (a) first harmonic (linear
response) and (b) second harmonic, for 1 eV gapped graphene and
$\tau_1 = \tau_2 = 5$ fs obtained using the time-dependent
calculation for $M=48$ cycles (black curves) and the
frequency-dependent method for infinitely periodic pulses.}
\label{fig:hcomp_spectral}
\end{figure}

\begin{figure}
\begin{center}
\includegraphics[width=0.49\textwidth]{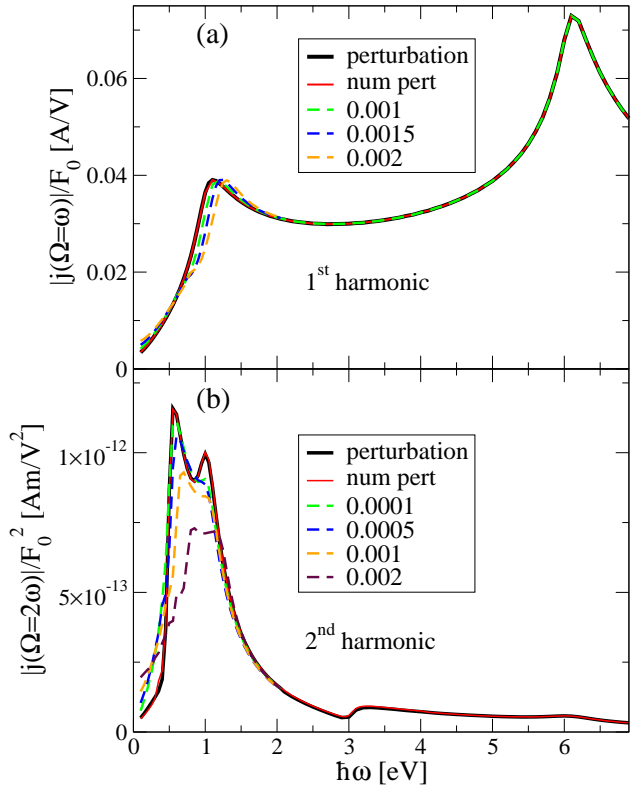}
\end{center}
\caption{The absolute value of the (a) first harmonic (linear
response) and (b) second harmonic, for 1 eV gapped graphene and
$\tau_1 = \tau_2 = 5$fs, obtained using the condition
\eqref{eq:n_condition} and compared to the perturbative result.
The numbers in the legends in panels (a) and (b) denote the peak
field strength (in a.u.) of the pulses used in the calculation.
The curve labelled 'num pert' denotes the numerical result that
compares best with the perturbation theory at peak fields (a)
$10^{-4}$ a.u., and (b) $10^{-5}$ a.u., see the text.}
\label{fig:1sthcomp}
\end{figure}

\begin{figure}
\begin{center}
\includegraphics[width=0.49\textwidth]{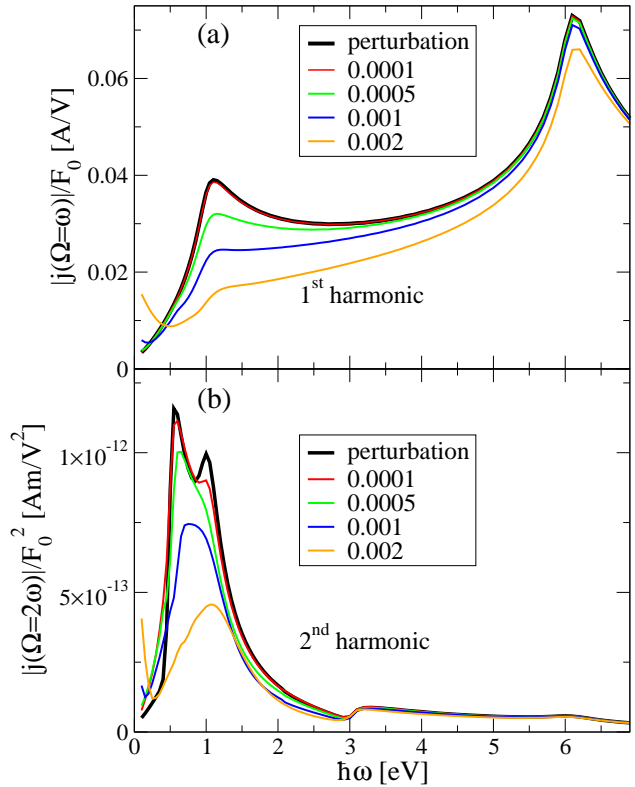}
\end{center}
\caption{The absolute value of the (a) first harmonic (linear
response) and (b) second harmonic, for 1 eV gapped graphene and
$\tau_1 = \tau_2 = 5$fs, obtained with full calculation and
compared to the perturbative result. The numbers in the legends in
panels (a) and (b) denote the peak field strength (in a.u.) of the
pulses used in the calculation.} \label{fig:2ndhcomp}
\end{figure}

The perturbative result in the time domain is obtained by
expanding $\rho_{cv}$ in orders of field strength as ($
\rho_{cv}^{(0)} = 0$ trivially)
\begin{equation}
\rho_{cv} ( \textbf{k},t) = \sum_{j=1}^{\infty} F_0^{j}
\rho_{cv}^{(j)} ( \textbf{k},t),  \label{eq:rho_pert_order}
\end{equation}
Inserting the condition of \eqref{eq:n_condition} into Eq.
\eqref{eq:eq_mot1} we obtain the following coupled system of
equations
\begin{eqnarray}
\frac{d \rho_{cv}^{(1)}}{dt} & = & -i \omega_{cv} \rho_{cv}^{(1)}
-i \frac{e}{\hbar} \Delta f_{vc} ( \textbf{k} ) \textbf{f} (t)
\cdot {\boldsymbol \xi}_{cv} - \frac{\rho_{cv}^{(1)}}{ \tau_1 }
\nonumber
\\
\frac{d \rho_{cv}^{(j)}}{dt} & = & -i \omega_{cv} \rho_{cv}^{(j)}
- \frac{\rho_{cv}^{(j)}}{ \tau_1 } -i \frac{e}{\hbar} \textbf{f}
(t) \cdot \left( {\boldsymbol \xi}_{cc} -{\boldsymbol \xi}_{vv}
\right) \rho_{cv}^{(j-1)} \nonumber \\ &+& \frac{e}{\hbar} \left(
\textbf{f} (t) \cdot {\boldsymbol \nabla}_{\textbf{k}} \right)
\rho_{cv}^{(j-1)} \quad \text{for} \quad j \ge 2 ,
\label{eq:eq_mot_pert}
\end{eqnarray}
where $\textbf{f}(t)=\textbf{F}(t) / F_0$ is the normalized field.
The first two coupled equations (for $j=1$ and $j=2$), that are
independent of the peak field strength, are solved putting $\tau_1
= \tau_2 = 5$fs to obtain the perturbative responses for the first
and the second harmonic in Figs.
\ref{fig:hcomp_spectral}-\ref{fig:2ndhcomp}.

We briefly review the features in the perturbative first and
second harmonic response. The absolute value of the linear
response (first harmonic) [Fig. \ref{fig:hcomp_spectral} (a)] has
peaks for photon energies corresponding to the gap $\Delta=1$ eV
and to the van Hove singularity \cite{vanHove}(M point - the point
where ${\boldsymbol \nabla}_\textbf{k} E_{c/v} ( \textbf{k} ) =0$)
at a photon energy of $2 \sqrt{( \Delta / 2)^2 + \gamma^2 | f (
\textbf{k} ) |^2}=6.2$ eV. The second harmonic response [Fig.
\ref{fig:hcomp_spectral} (b)], in addition to the peaks at the gap
and the van Hove singularity, should exhibit peaks at half of
these photon energies corresponding to two-photon transitions. It
is evident from Fig. \ref{fig:hcomp_spectral} that the peaks
corresponding to the van Hove singularity and to the half of this
frequency are very weak.

Next, we compare our perturbative solution obtained in the time
domain for $M=48$ cycles [black curves in Fig.
\ref{fig:hcomp_spectral}], with the corresponding solution for an
infinite periodic field, obtained using frequency-domain methods,
as done in Ref. \cite{TGP2015}, using $\tau_1 = \tau_2 =5$ fs (red
curves in Fig. \ref{fig:hcomp_spectral}). To compare directly, the
latter result is scaled (but not fitted) using appropriate factors
to the time-domain solution. This factor involves $N_p /2$ coming
from the Fourier transform [Eq. \eqref{eq:DFT}] and a factor
coming from the consideration of the limit of the type
$\lim_{\epsilon \to 0}  \exp (-\omega^2 / \epsilon^2) / ( \epsilon
\sqrt{\pi} ) = \delta ( \omega )$ for  the Fourier transform of
the envelope of the pulse [Eq. \eqref{eq:field}] (for the linear
response) and the square of the Fourier transform (for the second
harmonic). As evident from Fig. \ref{fig:hcomp_spectral}, the
agreement between the two types of solution is very good. There
are differences between the two methods at the peaks for the
second harmonic, whereas for photon energies away from the peaks
the agreement between the two methods is excellent.

The value of the time-domain perturbative solution is that
incorporates the finite pulse duration, so that a full numerical
solution for a finite number of cycles can be compared to it to
gauge the departure from the perturbative regime. In particular,
we compare the numerical solution with the solution in the
perturbative limit, that we also obtain numerically, for the
first, second and the third harmonic, and at photon energies
covering all significant features of the responses.


In the first set of results that we present, we perform numerical
calculation neglecting the time dependency of $n$, i.e., we use
Eqs. \eqref{eq:eq_mot1} and \eqref{eq:eq_mot2} keeping the
time-dependence of $n$ constant, equal to the initial value of
$n$, i.e.,
\begin{equation}
n( \textbf{k},t) = n( \textbf{k},-\infty)= f_v (\textbf{k}) -f_c
(\textbf{k}) = \Delta f_{vc} ( \textbf{k} ) .
\label{eq:n_condition}
\end{equation}
Then, Eq. \eqref{eq:eq_mot1} is solved with the above condition to
obtain the numerical result - this is essentially the solution in
the Keldysh approximation \cite{Keldysh1965}. Such an
approximation is used for semiclassical analysis of high-order
harmonic generation \cite{BrabecPRL2014,Brabec2015} in order to
simplify the theoretical analysis. The harmonic spectrum for the
first and the second harmonic response is divided by $F_0$ and
$F^2_0$, respectively. The results of these calculations are given
in Figs. \ref{fig:1sthcomp} (a) and \ref{fig:1sthcomp} (b),
respectively.

The second set of numerical results, given in Fig.
\ref{fig:2ndhcomp}, is obtained when the equations of motions
\eqref{eq:eq_mot1} and \eqref{eq:eq_mot2} are solved without
application of the condition \eqref{eq:n_condition}. In this way
both the effects of depletion of the band occupation $n$ and its
coupling with the coherences $\rho_{cv}$ are described. We refer
to this approach as the full calculation in the following.

We note that the perturbative first and the second harmonic
responses [Eqs. \eqref{eq:eq_mot_pert}], derived from the
equations of motion \eqref{eq:eq_mot1} and \eqref{eq:eq_mot2} with
or without the approximation for constant $n$ [Eq.
\eqref{eq:n_condition}] are identical. Namely, the intra current
from the second order in $n$ is zero because an odd function of
$\mathbf{k}$ is integrated, since $p_{cc} ( \mathbf{k} ) = -
p_{cc} (- \mathbf{k} ) $, $p_{vv} ( \mathbf{k} ) = - p_{vv} (-
\mathbf{k} ) $, $| \xi_{cv} ( \mathbf{k} )|^2 = | \xi_{cv} (-
\mathbf{k} ) |^2$, and $\Delta f_{vc} ( \textbf{k} ) = \Delta
f_{vc} ( - \textbf{k} ) $. Therefore the perturbative curve is
used in both Figs. \ref{fig:1sthcomp} and \ref{fig:2ndhcomp}.

In Figs. \ref{fig:1sthcomp} and \ref{fig:2ndhcomp} we compare the
numerically obtained responses with the perturbative responses. To
this end, we perform numerical calculations varying the peak field
strength until a certain harmonic response as a function of
frequency becomes 'frozen' for two consecutive field strengths.
This, 'frozen' curve for both first and second harmonic response
in Fig. \ref{fig:1sthcomp} is denoted as 'num pert'. For the first
harmonic this curve is obtained at a field strength of $10^{-4}$
a.u., whereas for the second harmonic that curve is obtained for a
field strength one order of magnitude smaller ($10^{-5}$ a.u.). As
evident from Fig. \ref{fig:1sthcomp}, the agreement of the
numerically extracted harmonic responses with the perturbative
responses is remarkable. Equally, in the case when we do not
invoke the approximation for $n=$const. [Fig. \ref{fig:2ndhcomp}]
we also obtain agreement with the perturbative result. We stress
that the results from the full calculation are not fitted to the
perturbative results, as done in Ref. \cite{Sipe2015}.

\section{Transition to the non-perturbative regime}

Gradually with the increase of the field strength,
non-perturbative features appear in the numerical responses,
starting at lower frequencies. This is visible in both cases:
calculations using condition \eqref{eq:n_condition} [Fig.
\ref{fig:1sthcomp}], and for the full calculation [Fig.
\ref{fig:2ndhcomp}]. In general, the peak field strengths at which
there is deviation from the perturbative results are smaller for
the second harmonic response than for the first harmonic response.
Next, for the first harmonic response, when using the full
calculation the deviation from the perturbative results
(calculated at equal peak field strengths) is larger compared to
the case when the condition of Eq. \eqref{eq:n_condition} is used;
compare Figs. \ref{fig:1sthcomp} (a) and \ref{fig:2ndhcomp} (a).
In case of the second harmonic response this difference is not
that large, however, it is non-negligible [Figs.
\ref{fig:1sthcomp} (b) and \ref{fig:2ndhcomp} (b)]. This is
striking since in all our numerical calculations, during the time
evolution, the depletion of $n$ is at most 1\% at the largest peak
field strength used. This exposes the inadequacy of the
approximation of Eq. \eqref{eq:n_condition} even at very small
field strengths - in the discussion below we therefore use results
obtained using the full calculation. Finally, while for the first
harmonic the yield essentially decreases preserving the shape as
the peak field strength increases, for the second harmonic
response the modification is not only a decrease in magnitude but
also the shape of the response is changed in such a way that the
peaks at low energy (0.5 and 1 eV) merge into one rounded peak
[Fig. \ref{fig:2ndhcomp} (b)]. We note that in the linear regime
the ratio of the peak of the field generated by the $n$-th
harmonic to the incident peak field $F_0$ is approximately equal
to the ratio $|j ( \Omega = n \omega ) | / |j ( \Omega = \omega )
|$. For example for the second harmonic, in the worst case when
$F_0 = 0.002$ a.u. is used, this ratio is of the order of
$10^{-2}$.

\begin{figure}
\begin{center}
\includegraphics[width=0.49\textwidth]{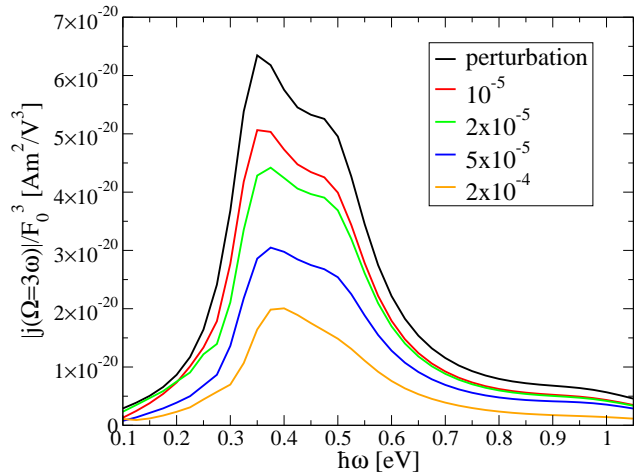}
\end{center}
\caption{The absolute value of the third harmonic for 1 eV gapped
graphene, $\tau_1 = \tau_2 =5$ fs, obtained at different peak
field strengths and compared to the perturbative result. The
numbers in the legend denote the peak field strength (in a.u.) of
the $M=48$ cycle pulses used in the calculation.}
\label{fig:3rdhcomp}
\end{figure}

The departure from the pertubative regime can be illustrated for
the third harmonic as well. In contrast to the first and the
second harmonic, the perturbative limit for the third harmonic
contains not only contribution from the inter part of the current,
but also from the intra part of the current. Therefore the
equations \eqref{eq:eq_mot_pert} are inadequate to describe the
pertubative third harmonic generation and should be completed by
adding equations for the coefficients $n^{(j)}$, $j \le 3$, of the
perturbative expansion of $n$, i.e.,
\begin{equation}
n  ( \textbf{k},t ) = \sum_{j=1}^{\infty} F_0^{j} n^{(j)} (
\textbf{k},t),  \label{eq:n_pert_order}
\end{equation}
Then we insert the above expansion and the perturbative expansion
of $\rho$ in Eq. \eqref{eq:rho_pert_order} in the equations of
motion \eqref{eq:eq_mot1} and \eqref{eq:eq_mot2}. This procedure
results in adding the following equations for $n^{(2)}$ and
$n^{(3)}$ ($n^{(0)} = \Delta f_{vc} ( \textbf{k} ) $ and $n^{(1)}
= 0$)
\begin{eqnarray}
\frac{d n^{(2)}}{dt} & = & 2 i \frac{e}{\hbar} \mathbf{f} (t)
\cdot \left(  {\boldsymbol \xi}_{cv} {\rho^{(1)}_{cv}}^* -
{\boldsymbol
\xi}^*_{cv} {\rho^{(1)}_{cv}} \right) - \frac{n^{(2)}}{\tau_2} \nonumber \\
\frac{d n^{(3)}}{dt} & = & 2 i \frac{e}{\hbar} \mathbf{f} (t)
\cdot \left(  {\boldsymbol \xi}_{cv} {\rho^{(2)}_{cv}}^* -
{\boldsymbol \xi}^*_{cv} {\rho^{(2)}_{cv}} \right)  -
\frac{n^{(3)}}{\tau_2} \nonumber \\ & + & \frac{e}{\hbar}
\mathbf{f} (t) \cdot ( {\boldsymbol \nabla}_{\mathbf{k}} n^{(2)})
\label{eq:n_pert}
\end{eqnarray}
to the system of equations \eqref{eq:eq_mot_pert} and modifying
the equation for $\rho^{(3)}_{cv}$ as
\begin{eqnarray}
\frac{d \rho_{cv}^{(3)}}{dt} & = & -i \omega_{cv} \rho_{cv}^{(3)}
- \frac{\rho_{cv}^{(3)}}{ \tau_1 } -i \frac{e}{\hbar} \textbf{f}
(t) \cdot \left( {\boldsymbol \xi}_{cc} -{\boldsymbol \xi}_{vv}
\right) \rho_{cv}^{(2)} \nonumber \\ &+& \frac{e}{\hbar} \left(
\textbf{f} (t) \cdot {\boldsymbol \nabla}_{\textbf{k}} \right)
\rho_{cv}^{(2)} - i \frac{e}{\hbar} \mathbf{f} (t) \cdot
{\boldsymbol \xi}_{cv} n^{(2)} .\label{eq:rhocv3}
\end{eqnarray}
Using these equations, the perturbative third harmonic is
obtained. In Fig. \ref{fig:3rdhcomp}, this perturbative result
(with $\tau_1 = \tau_2 = 5$fs) is compared to the full numerical
calculation at different peak field strengths for photon energies
up to 1 eV, as for higher photon energies the response falls off
rapidly to zero. The perturbative curve has peaks at photon
energies corresponding to one third and one half of the gap. As
the field increases, the height of the harmonic decreases and the
peaks merge into one broad peak. The discrepancy between the full
calculation and the perturbative result starts here at lower peak
field strengths (at least as small as $10^{-6}$ a.u.) as compared
to both the first and the second harmonic response.

\begin{figure}
\begin{center}
\includegraphics[width=0.49\textwidth]{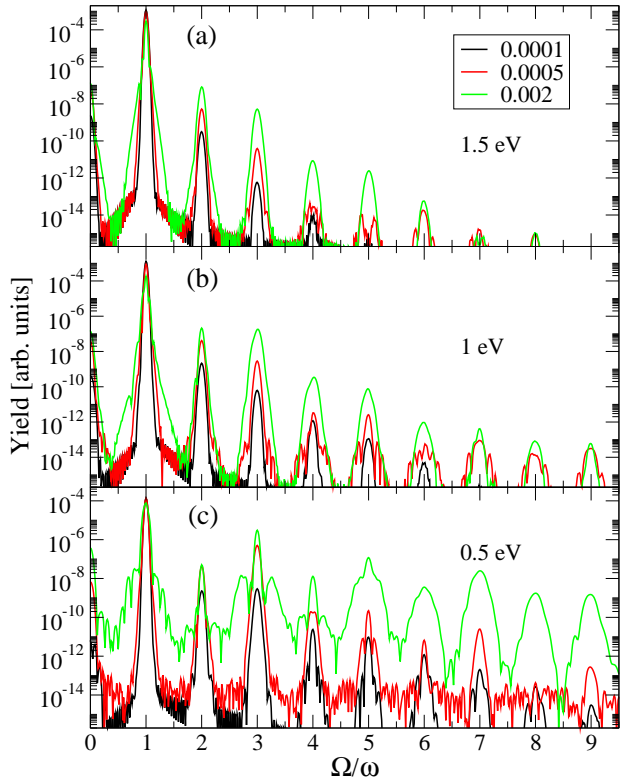}
\end{center}
\caption{Harmonic spectra (divided by peak field strength squared)
for different peak fields (given in the legend in atomic units) at
(a) $\hbar \omega=1.5$ eV, (b) $\hbar \omega = 1$ eV, and (c)
$\hbar \omega = 0.5$ eV photon energy. The yield is proportional
to $ |j( \Omega ) |^2 / F_0^2$.} \label{fig:spectra_all}
\end{figure}

Figure \ref{fig:spectra_all} shows what happens to the harmonic
spectra after the field strength becomes large enough and/or the
incident photon energy becomes small enough so that perturbation
theory breaks down. We note that the harmonic spectra depicted in
Fig. \ref{fig:spectra_all} are divided by the square of the peak
field strength so that the first harmonic is at comparable height
for different field strengths. In Fig. \ref{fig:spectra_all} (a)
the situation for a photon energy of 1.5 eV is depicted. At the
perturbative field strength of $10^{-4}$ a.u. the height of the
higher harmonics rapidly falls off. This is also true for the next
larger peak field strength in Fig. \ref{fig:spectra_all} (a).
However, for the highest peak field strength, the fall-off is not
so rapid and pairs of adjacent harmonics (2nd and 3rd, 4th and
5th) tend to almost level up in height. At a lower photon energy
of 1 eV [Fig. \ref{fig:spectra_all} (b)] and at the largest peak
field strength the beginning of a plateau, known to be typical for
atoms and molecules \cite{corkum_plasma_1993,Lewenstein1994}, is
visible. For the lowest photon energy depicted [0.5 eV in Fig.
\ref{fig:spectra_all} (c)], the harmonic spectrum forms a
pronounced plateau for the two largest peak field strengths. The
number of harmonics forming the plateau is roughly proportional to
the peak field strength. This is in qualitative agreement with a
semiclassical analysis for the harmonic cutoff \cite{Brabec2015},
where it was predicted that it is proportional to $F_0 / \omega$.
After the departure from the perturbative regime, due to the
increase of this factor, the harmonic peaks start forming a
plateau, which is a signature of non-perturbative dynamics.


\begin{figure}
\begin{center}
\begin{tabular}{c}
\includegraphics[width=0.49\textwidth]{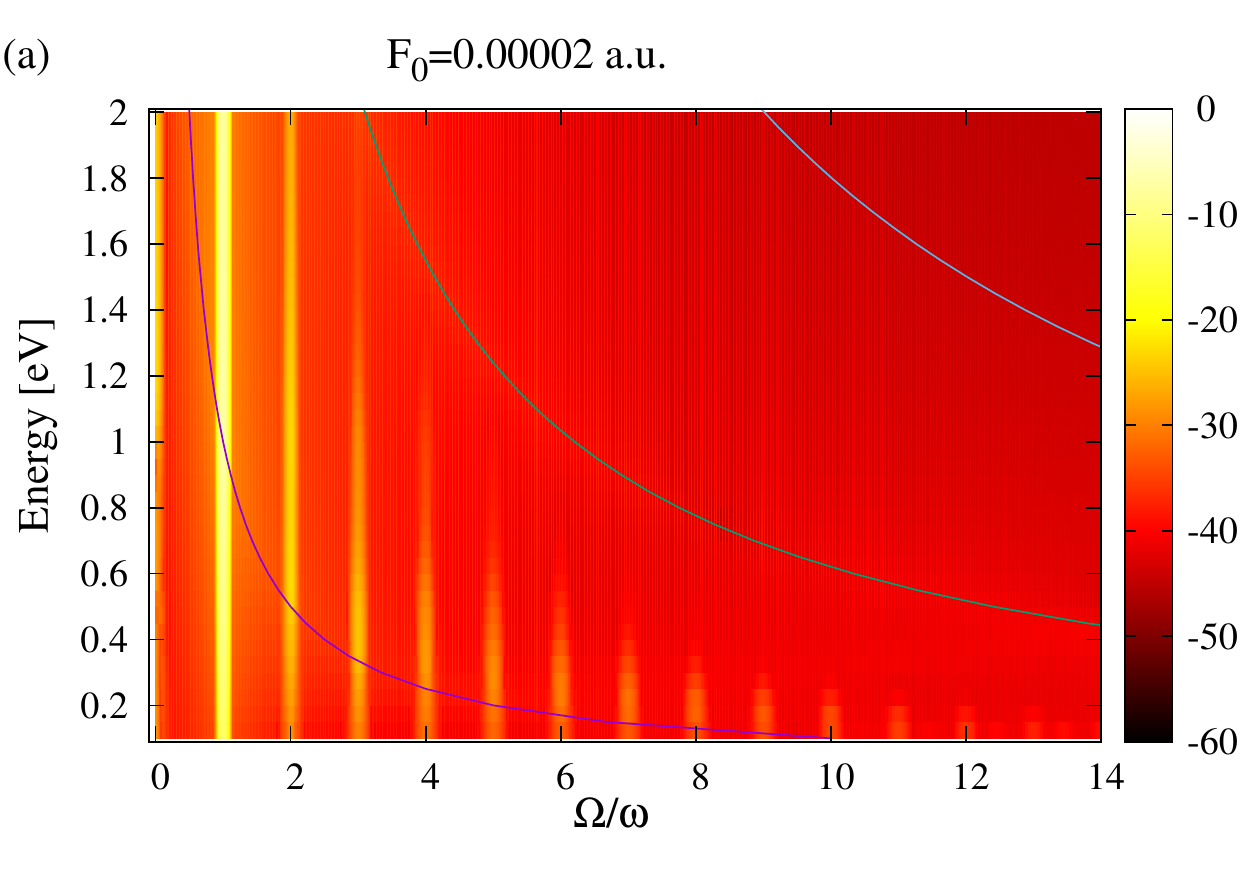} \\
\includegraphics[width=0.49\textwidth]{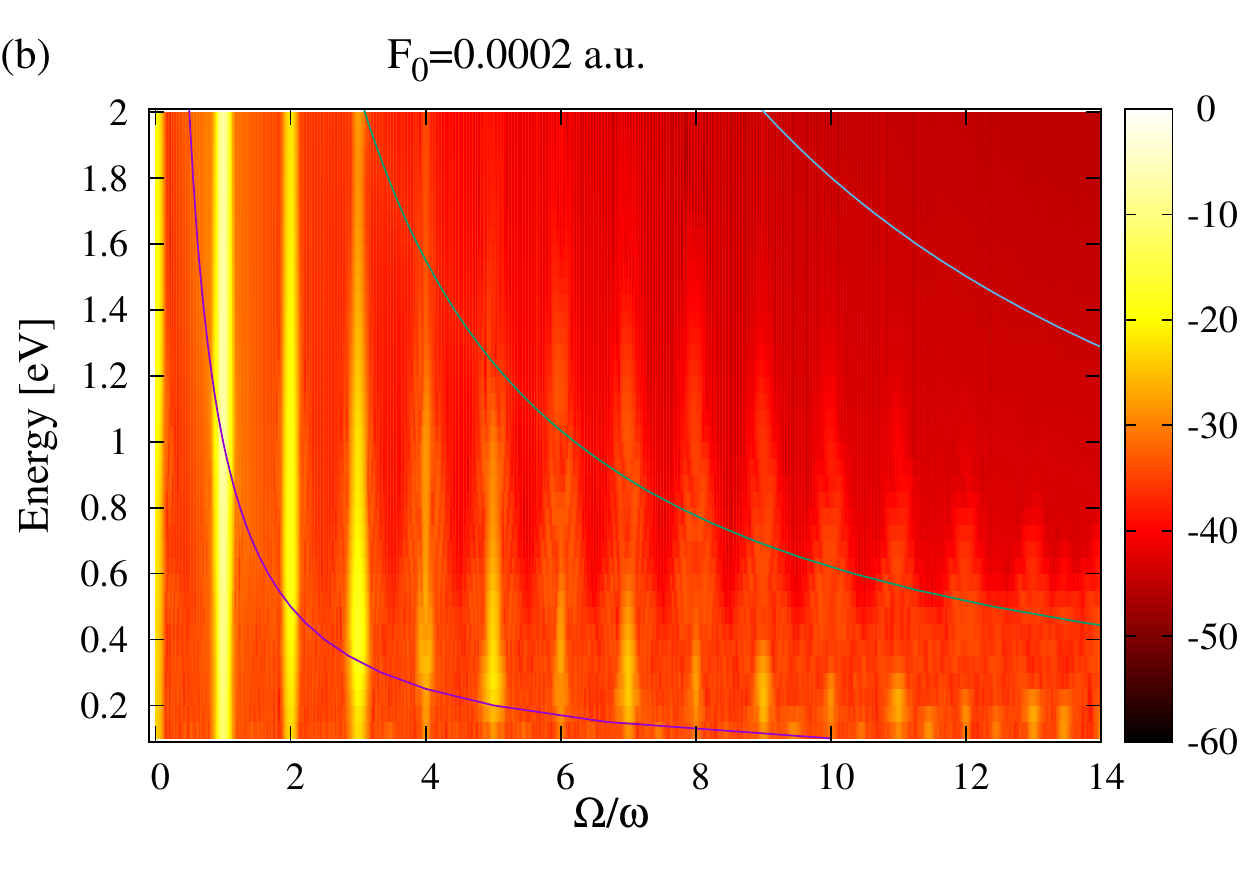} \\
\includegraphics[width=0.49\textwidth]{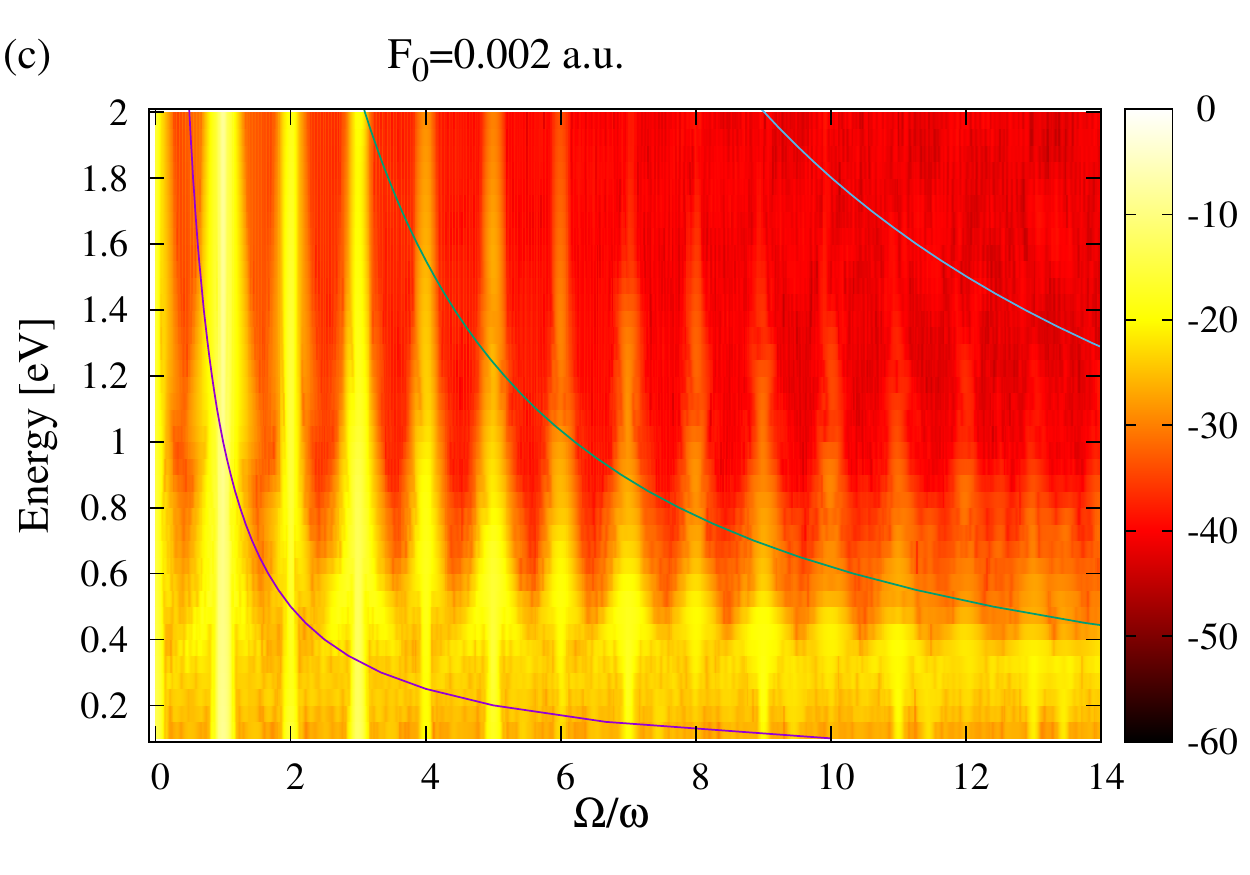}
\end{tabular}
\end{center}

\caption{Harmonic spectra (divided by peak field strength squared)
at different photon energies of the driving field on color
logarithmic scale in arbitrary units. (a) spectra for a peak field
strength of $2 \times 10^{-5}$ a.u. (close to the perturbation
regime), (b) spectra for a peak field strength of $2 \times
10^{-4}$ a.u., and (c) spectra for a peak field strength of $2
\times 10^{-3}$ a.u. (deeply in the non-perturbative regime). The
lines in the color plots denote the borders defining how many
harmonics fit (in order from left to right on the figure) at the 1
eV gap (K point), at the 6.2 eV gap at van Hove singularity (M
point), and at the maximum gap of 18.03 eV ($\Gamma$ point).}
\label{fig:differential}
\end{figure}

The transition to the non-perturbative regime is also illustrated
in Fig. \ref{fig:differential}, where harmonic spectra are given
as two-dimensional plots of the harmonic order and the photon
energy of the driving field in the interval from 0.1 to 2 eV, and
for different field strengths. We present this figure to
illustrate the growth of the harmonics at different photon
energies as the peak field strength increases. A single horizontal
line in Fig. \ref{fig:differential} contains a harmonic spectrum
of the type presented in Fig. \ref{fig:spectra_all}. To
qualitatively estimate the progression of harmonics as the peak
field strength increases, the two-dimensional space (harmonic
order, photon energy) is divided by three curves, corresponding to
the borders of how many harmonics fit in (in order from left to
right in Fig. \ref{fig:differential}) the 1eV gap, the gap
corresponding to the van Hove singularity (6.2 eV), and the
maximum gap (18.03 eV) in our two-band model.

The spectra for the smallest field strength [Fig.
\ref{fig:differential} (a)] contain well-pronounced harmonics
which drop off in the (harmonic order, photon energy) region
bounded by the curves corresponding to the gap and van Hove
singularity, see the caption of Fig. \ref{fig:differential}. The
harmonic peaks for the next larger peak strength [Fig.
\ref{fig:differential} (b)] drop off around the van Hove
singularity curve. Lastly, the harmonics at the largest peak field
strength considered [Fig. \ref{fig:differential} (c)] drop off in
the region bounded by the curves corresponding to the van Hove
singularity and the maximum gap. The curve corresponding to the
maximum gap is in fact the limit for harmonic generation in the
present two-band model - no well-formed harmonic at any field
strengths is situated to the right of this curve. We note that at
the energy range occupied by the harmonics of higher orders the
contribution from other bands may not be ignored. Here, however,
we only consider the non-perturbative limit within the two band
model.

\begin{figure}
\begin{center}
\includegraphics[width=0.49\textwidth]{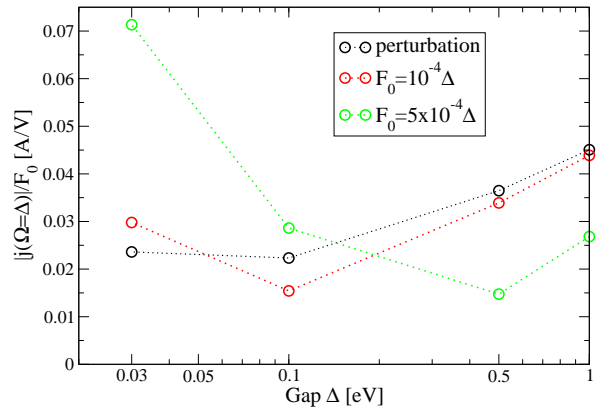}
\end{center}
\caption{First harmonic at the photon energy equal to the gap
($\hbar \omega = \Delta$). The peak field strengths (scaled to the
gap) are given in a.u. in the legend. To obtain the actual field
strength in a.u. used for a given gap, $\Delta$ in the legend
should be given in eV.} \label{fig:h1_pert}
\end{figure}

Finally, we consider the gap dependence. For simplicity, we focus
at photon energies corresponding to the gap ($\hbar \omega =
\Delta$), where the major part of the first order response is
located and where the deviation from the perturbative result is
more pronounced. We aim to compare different gaps for field
strengths that result in comparable values of the response. A
possible scaling for the field strength emerges by considering
that the leading order of the dependence of the dipole matrix
element ${\boldsymbol \xi}_{cv}$ is $\Delta^{-1}$. Assuming that
this term is dominant in the differential equations of motion
[Eqs. \ref{eq:eq_mot1} and \ref{eq:eq_mot2}], when changing the
gap $\Delta$, a field $ c \Delta F_0 $, where $c$ is a constant,
will give roughly, but not exactly, the same response. To limit
the total duration of the numerical time propagation, we consider
pulses with $M=12$ which are long enough to be free from few-cycle
effects. Similarly to Ref. \cite{Cheng2015}, we use an asymmetric
decoherence times, with $\tau_1 = 10$ fs and $\tau_2 = 1$ ps. The
results of the calculations are shown in Fig. \ref{fig:h1_pert}.
The scaled first harmonics in the figure are of the same order of
magnitude, which justifies the scaling of the field strength.
Moreover, as the gap decreases the perturbative result (obtained
using Eqs. \eqref{eq:eq_mot_pert}] becomes more flat, reflecting
the fact that the term ${\boldsymbol \xi}_{cv}$ becomes more
dominant in the equations of motion. From the other curves, the
rough scaling of the peak field strength at which the perturbation
theory breaks down can be deduced. Namely, the curve corresponding
to the peak field strength that gives almost perturbative result
at 1eV gap (the curve labelled with $F_0 = 10^{-4} \Delta$ in Fig.
\ref{fig:h1_pert}) becomes a bit more non-perturbative as the gap
decreases. Hence, for the first harmonic, it is safe to assume
that if there is a deviation between the perturbative result and
the full calculation at 1 eV , this deviation will be even larger
for the gaps at an equivalent scaled peak electric field. For a
larger field strength (the curve labelled with $F_0 = 5 \times
10^{-4} \Delta$ in Fig. \ref{fig:h1_pert}), the result is already
deep in the non-perturbative regime for a gap of 1eV, and at
smaller gaps it enters even deeper in the non-perturbative regime.

\begin{figure}
\begin{center}
\includegraphics[width=0.49\textwidth]{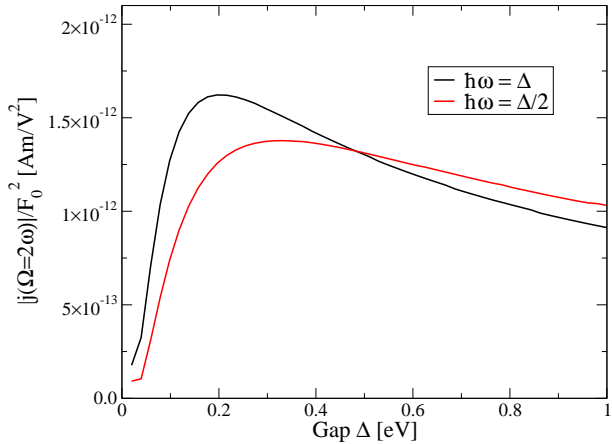}
\end{center}
\caption{Second harmonic at the photon energy equal to the gap
($\hbar \omega = \Delta$) and at photon energy equal to the half
of the gap ($\hbar \omega = \Delta/2$), obtained using the
frequency-domain method of Ref. \cite{TGP2015}, $\tau_1 =\tau_2 =
5$ fs, and $6000 \times 6000$ grid in $\mathbf{k}$-space.}
\label{fig:h2spectral}
\end{figure}

In closing, we consider the gap dependence of the second harmonic
in Fig. \ref{fig:h2spectral}. This is interesting since in the
limit of zero gap the second-order harmonic vanishes. To
investigate this limit it is easier to use the frequency-domain
method than the time-dependent one since as the gap decreases
larger grids in $\mathbf{k}$-space should be taken and the pulses
should be propagated for longer times, which becomes prohibitively
time-consuming. Therefore in Fig. \ref{fig:h2spectral}, where we
plot the second harmonic at a driving photon frequency equal to
the gap and to the half of the gap, respectively, we used the
frequency-domain method of Ref. \cite{TGP2015}. We applied the
same scaling factors as for the results from the frequency-domain
method presented in Fig. \ref{fig:hcomp_spectral} (b). As the gap
decreases, the height of the second harmonic first increases,
reaching a peak at approximately at 0.2 eV for the case of $\hbar
\omega = \Delta$ and approximately at 0.3 eV for the case of
$\hbar \omega = \Delta/2$, and then falls towards zero. This is so
since as the gap starts decreasing, (i) the energies in the
denominators of the expression for the second order conductivity
[Eq. (27) in Ref. \cite{TGP2015}] become small, but also (ii) the
numerators of the same expression become smaller as the
centrosymmetric limit is approached. Eventually, the numerator
wins and the second harmonic current goes to zero. The same was
observed for the second harmonic in carbon nanotubes
\cite{PedersenNTube2009} as the radius of the tube increases and
the tube approaches the planar graphene limit.

\section{Conclusions and outlook}

We have explored the limit of perturbative harmonic response,
which is usually considered for infinitely periodic pulses, in the
time domain, and demonstrated excellent agreement between the
numerical calculation and perturbation theory for low laser
intensity over the interval of photon energies that includes all
features in the spectrum. The numerical method for perturbative
harmonic responses is especially well-suited to obtain not only
the first few harmonics, but also high-order harmonics for
realistic, finite-duration pulses.

Comparing with the full non-perturbative calculations, we conclude
that the harmonic response starts to deviate from the perturbative
harmonic response at relatively low field strengths. The
calculation performed for constant difference in the occupation of
the valence and the conduction band fails to reproduce the correct
point of departure from the perturbative result even for the first
harmonic, which exposes its weakness.

Finally, we have illustrated the transition from the perturbative
to the non-perturbative regime in the harmonic spectra. For each
harmonic, the breakdown of perturbation theory occurs at different
field strength, which is smaller for the second harmonic than for
the first harmonic. For the third harmonic the perturbation theory
breaks down at even smaller field strengths. Increasing the field
strength further, the harmonics start forming the typical HHG
plateau, well-studied in the strong-field physics for atoms and
molecules. In contrast to atoms and molecules, the plateau cut-off
is here limited by the maximum gap since the analysis was
performed in a two-band approximation. At the end, we have
considered the gap dependence for the linear response using simple
scaling, and illustrated the transition to the non-perturbative
regime.

As strong-field physics with its intense near-infrared laser
pulses of femtosecond duration is extended from atoms and
molecules to condensed matter systems \cite{Ghimire2011}, and with
the advent of high-harmonic spectroscopy for solids
\cite{Gouliemakis2015}, the development of theory that is
explicitly time-dependent and capable of dealing with the
laser-matter interaction in a non-perturbative manner is
essential. Here we provided a candidate for such a theory which,
in this work, was validated by comparison with the results of
perturbation theory. The formulation can be extended to multiple
bands, combination of pulses, other materials, and to include the
Coulomb interaction. It is probably in these contexts that the
coherence properties of the laser light and the ability to perform
pump-probe experiments and simulations will show its full
potential for gaining time-resolved insight in ultrafast dynamics
in solids.

\section*{Acknowledgements}

This work was supported by the Villum Kann Rasmussen (VKR) center
of excellence, QUSCOPE. The numerical results were obtained at the
Centre for Scientific Computing, Aarhus.

\section*{Appendix: Dipole and momentum matrix elements}

\subsection{Dipole matrix elements ($ {\boldsymbol \xi}_{nm} $)}

The eigenvectors of the Hamiltonian $\hat{\textbf{H}}_0$ of Eq.
\eqref{eq:GG} are

\begin{equation}
\left| n \right> = \frac{1}{\sqrt{2}}  \left( {\begin{array}{c}
    \sqrt{(E_n + \Delta/2)/E_n}
   \\ \pm e^{-i \phi ( \mathbf{k} ) } \sqrt{(E_n - \Delta/2)/E_n}
\end{array} }
   \right) , \label{eq:eigen_plus}
\end{equation}
where $ \left| n \right>$ denotes either the states in the
conduction ($\left| c \right>$) or the valence band ($\left| v
\right>$), $E_n$ denotes either $E_{c}$ or $E_{v}$, '$\pm$' is '+'
for the conduction and '-' for the valence band, respectively, and

\begin{equation}
\phi ( \mathbf{k} ) = \text{Arg} [ f ( \mathbf{k} ) ],
\label{eq:arg}
\end{equation}
with $f ( \mathbf{k} )$ given in Eq. \eqref{eq:fk}.

The dipole matrix elements ${\boldsymbol \xi}_{cv}$ and
${\boldsymbol \xi}_{cc} - {\boldsymbol \xi}_{vv}$, used in the
main text, are obtained by direct calculation, i.e., by
calculating $\left< n \right| i {\boldsymbol \nabla }_{\mathbf{k}}
\left| m \right>$, $n,m = c,v$. They are explicitly given by

\begin{widetext}

\begin{equation}
\text{Re} \left\{ {\boldsymbol \xi}_{cv} \right\} = \frac{a
\gamma}{2 E_c \left| f ( \textbf{k} ) \right|} \left[
\frac{1}{\sqrt{3}} \left( \cos ( a k_x \sqrt{3}/2 ) \cos ( a k_y
/2 ) - \cos ( a k_y ) \right) \mathbf{e}_x + \sin ( a k_x
\sqrt{3}/2 ) \sin ( a k_y /2 ) \mathbf{e}_y \right],
\end{equation}

\begin{equation}
\text{Im} \left\{ {\boldsymbol \xi}_{cv} \right\} = \frac{a \Delta
\gamma}{4 E^2_c \left| f ( \textbf{k} ) \right|} \left[ \sqrt{3}
\sin ( a k_x \sqrt{3}/2 ) \cos ( a k_y /2 ) \mathbf{e}_x + \left(
\cos ( a k_x \sqrt{3}/2 ) \sin ( a k_y /2 ) + \sin ( a k_y )
\right) \mathbf{e}_y \right],
\end{equation}
and
\begin{equation}
{\boldsymbol \xi}_{cc} - {\boldsymbol \xi}_{vv}  = - \frac{a
\Delta}{2 \sqrt{3} E_c \left| f ( \textbf{k} ) \right|^2} \left[
 \left( \cos ( a k_x \sqrt{3}/2 ) \cos ( a k_y
/2 ) - \cos ( a k_y ) \right) \mathbf{e}_x + \sin ( a k_x
\sqrt{3}/2 ) \sin ( a k_y /2 ) \mathbf{e}_y \right] .
\end{equation}

\subsection{Momentum matrix elements ($\mathbf{p}_{nm}$)}

The diagonal momentum matrix elements are obtained as $
\textbf{p}_{nn} = \dfrac{m_e }{\hbar} \dfrac{\partial E_n
}{\partial \textbf{k}}$ yielding

\begin{equation}
\textbf{p}_{cc} = - \frac{m_e}{\hbar} \frac{a \gamma^2}{ E_c }
\left[ \sqrt{3} \sin ( a k_x \sqrt{3}/2 ) \cos ( a k_y /2 )
 \mathbf{e}_x +   \left( \cos ( a k_x \sqrt{3}/2 )
\sin ( a k_y /2 ) + \sin ( a k_y ) \right) \mathbf{e}_y \right],
\quad \textbf{p}_{vv} = - \textbf{p}_{cc} .
\end{equation}

The off-diagonal matrix element $\textbf{p}_{cv}$ is simply
obtained using

\begin{equation}
\mathbf{p}_{cv} = i \frac{m_e}{\hbar} \left( E_c - E_v \right)
{\boldsymbol \xi}_{cv} . \label{eq:mom_off_diag}
\end{equation}

\end{widetext}

\end{document}